\documentclass{article}

\usepackage[margin=2.5cm]{geometry}
\usepackage[utf8]{inputenc} % Accents
\usepackage{graphicx} % Figures
\usepackage{url} % To formally define an URL
\usepackage[pdftex]{hyperref} % To create hyperlinks in PDF
\usepackage{color} % For colored parts
\usepackage{verbatim} % For comments
\usepackage{textcomp} % For TM
\usepackage{subcaption}
\usepackage{rotating}

\hypersetup{
colorlinks,%
citecolor=black,%
filecolor=black,%
linkcolor=black,%
urlcolor=black
}

\title{BEAT: An Open-Source Web-Based Open-Science Platform}

\author{Andr\'{e} Anjos, Laurent El-Shafey and S\'{e}bastien Marcel \\
\small\texttt{\{andre.anjos,laurent.el-shafey,sebastien.marcel\}@idiap.ch} \\
\small Idiap Research Institute, rue Marconi, 19 \\
\small Centre du Parc, 1920, Martigny, VS, Switzerland}

\begin{document}
\maketitle
\begin{abstract}

With the increased interest in computational sciences, machine learning (ML),
pattern recognition (PR) and big data, governmental agencies, academia and
manufacturers are overwhelmed by the constant influx of new algorithms and
techniques promising improved performance, generalization and robustness.
Sadly, result reproducibility is often an overlooked feature accompanying
original research publications, competitions and benchmark evaluations. The
main reasons behind such a gap arise from natural complications in research
and development in this area: the distribution of data may be a sensitive
issue; software frameworks are difficult to install and maintain; Test
protocols may involve a potentially large set of intricate steps which are
difficult to handle. Given the raising complexity of research challenges and
the constant increase in data volume, the conditions for achieving reproducible
research in the domain are also increasingly difficult to meet.

To bridge this gap, we built an open platform for
research in computational sciences related to pattern recognition and machine
learning, to help on the development, reproducibility and certification of
results obtained in the field. By making use of such a system, academic,
governmental or industrial organizations enable users to easily and socially
develop processing toolchains, re-use data, algorithms, workflows and compare
results from distinct algorithms and/or parameterizations with minimal effort.
This article presents such a platform and discusses some of its key features,
uses and limitations. We overview a currently operational prototype and provide
design insights.


\end{abstract}

\section{Introduction}

One of the key aspects of modern computer science research lies in the use of
personal computers (PCs) either for the simulation of known phenomena or for
the evaluation of data collected from natural observations. Mashups of these
data, organized in tables and figures are attached to textual descriptions
leading to scientific publications. Frequently, data sets, code and actionable
software leading to results are excluded upon recording and preservation of
articles. This situation slows down potential scientific development in at
least two major aspects: (1) re-using ideas from different sources normally
implies the re-development of software leading to original results and (2) the
reviewing process of candidate ideas is based on trust rather than on hard,
verifiable evidence \cite{price:1986}.

The need and benefits for reproducibility in computational science was already
recognized by academia~\cite{nature:2016,jasny:2011,vandewalle:2012} and
industry~\cite{executable-paper}, though concrete actions to overcome inherent
difficulties are yet to appear into \textit{de facto} standards in this field.
For example, it has been shown by MIT researchers~\cite{mit-generator} that the
reviewing process that determines article acceptance in some conferences may by
tricked by publications with machine generated content.

While scientific articles normally incorporate a stage of certification
referred as peer-reviewing, the very same software frameworks and data leading
to the stated conclusions, \textit{when available}, are considered as a bonus
and dismissed unreviewed. Even if knowledgable reviewers can predict when
written material is insufficiently discussed or poorly argued, one must also
consider the hypothesis of rich arguments being coupled to poorly executed
software implementations and data quality ending up in misleading conclusions,
which do not translate in scientific development. It is a fact that correct and
repeatable execution of software over data does not guarantee a good trend
either, but less so does just an article. Only by coupling scientific reports,
software and data to the extensibility and reuse which is required to verify
evidence, can society achieve a faster and steady pace of development.
Reproducibility and certification, in this context, should become a must for
the future of articifial intelligence rather than a mere bonus.

Publication of research results is not the sole place where advances are
needed. The conception of ideas, their embodiment in the form of computer
routines and experiments, as well as the actual reviewing process could also
benefit from technological advances in computer cloud infrastructures,
programming tools (e.g. Web 2.0) and social networking. Albeit limiting,
technological challenges do not hold exclusivity in irreproducibility. Many
research domains such as those related to medical, biometrics and forensics
applications also face legal barriers. Data used in these domains should be
handled according to stringent law requirements related to human rights for
privacy, which poses obstacles to reproducibility, but also knowledge sharing.

\subsection{The \textit{Status Quo}: what can we do better?}

The idea behind a platform for the evaluation of reproducible machine learning
and pattern recognition algorithms is not new. Software-based frameworks
currently exist in different implementation languages and to attend different
purposes. The current main trend seems to be biased towards the creation of web
services that ease the management of challenges in machine learning and pattern
recognition~\cite{codalab, kaggle} instead of run-yourself software solutions
which were very popular in the last decade. Web services can offer convenient
access through different types of devices (computers, tablets and mobile
phones), while requiring only a compatible web-browser to be installed on the
user machine. With the advent of modern web programming techniques and useful
libraries, there is virtually nothing one cannot do through a browser window.

One of the key issues with leading platforms on this market is that of data
sharing and privacy. At the same time reproducibility in computational science
calls for open data access, certain research domains must respect privacy
considerations when sharing data. With new EU privacy law requirements well on
the way~\cite{eu-law-reform} and matching agreements being reached with other
leading countries, personal data transfer must respect formal conditioning and
safe keeping - for example, biometric data may be accessible only via end-user
license agreements which, typically, disallow copying outside institutional
premises. Such a trend will directly influence how research is able to access
and share data which, in turn, will impact the adoption of existing solutions.
In practice, platforms tackling research reproducibility must incorporate
privacy by
design\footnote{\url{https://en.wikipedia.org/wiki/Privacy_by_design}} (PbD) on
their blue prints. PbD can be beneficial to key players in academic domains in
which data is not easily transferable, but also to improve the relationship
between those, industry and governamental agencies which are also key players
in research.

When one talks about research in machine learning and pattern recognition, they
must not forget difficulties related to the implementation of its software
building blocks, required for the needed repetitive testing, evaluation and
performance tuning leading to discovery and reading material. Each of those
blocks is implemented over and over using fashionable languages and paradigms
through time, existing in a format which becomes outdated as new fashions and
paradigms appear. At a point in time, the FORTRAN language was considered the
\textit{de facto} scientific programming tool. After that Matlab and nowadays a
myriad of options exist to encode knowledge in this domain. You must have asked
yourself many times \textit{``Which to pick?"}. Because no right answer is on
the horizon, solutions must also take into consideration the hybrid nature in
this research domain. Building workflows require tools from a variety of
languages to co-exist through time to build the perfect re-usable machinery.

In the remainder of this article, we introduce the BEAT
platform,\footnote{Operational at
\url{https://www.beat-eu.org/platform/}.} a PbD-built architecture to take on
these issues: social development, hybrid algorithm re-use, open-sourcing and
confidentiality, providing a new paradigm for the development and evaluation
of pattern recognition tools. We present the platform design in
Section~\ref{sec:design}, outlining its main components and core technology. In
Section~\ref{sec:use} we examplify how it can be use to address data-driven
problems in computational science through different use-cases in education,
challenge preparation and industry-academia relationship. Finally, at
Section~\ref{sec:conclusions}, we conclude the article with a summary of
platform limitations and a look into its future.

\section{The BEAT Platform}
\label{sec:design}

BEAT is a pan-european project composed of both academic and industrial
partners in which one of the goals was the design and development of a free,
open-source,\footnote{Source-code: \url{https://gitlab.idiap.ch/beat/}} online
web-based platform for the development and certification of reproducible
software-based machine learning (ML) and pattern recognition (PR) experiments.
The main intent behind the platform is to establish a framework for the
certification and performance analysis of such systems while still respecting
privacy and confidentiality of built-in data and user contributions. The
framework, as per definition, is task-independent, being adaptable to different
problem domains and evaluation scenarios. At the conceptual phase, the platform
was bound to support a number of use-cases which we try to summarize:

\begin{itemize}
  \item Benchmarking of ML and PR systems and components: users should be able
    to program and execute full systems so as to identify performance and
    computing requirements for complete toolchains or individual components;
  \item Comparative evaluation: it should be possible to run challenges and
    competitions on the platform as it is the case in similar systems such as
    Kaggle \cite{kaggle};
  \item Certification of ML and PR systems: the platform should be able to
    attest on the operation and performance of experiments so as to support the
    work of certification agencies or publication claims;
  \item Educational resource: the platform shall be usable as an educational
    resource for transmitting know-how about ML and PR applications. It should
    be possible to set-up interest groups that share work assignments such as
    in a teacher-student relationship.
\end{itemize}

\subsection{Application Breakdown}

\begin{figure}[htpb]
  \centering
  \includegraphics[width=0.6\textwidth]{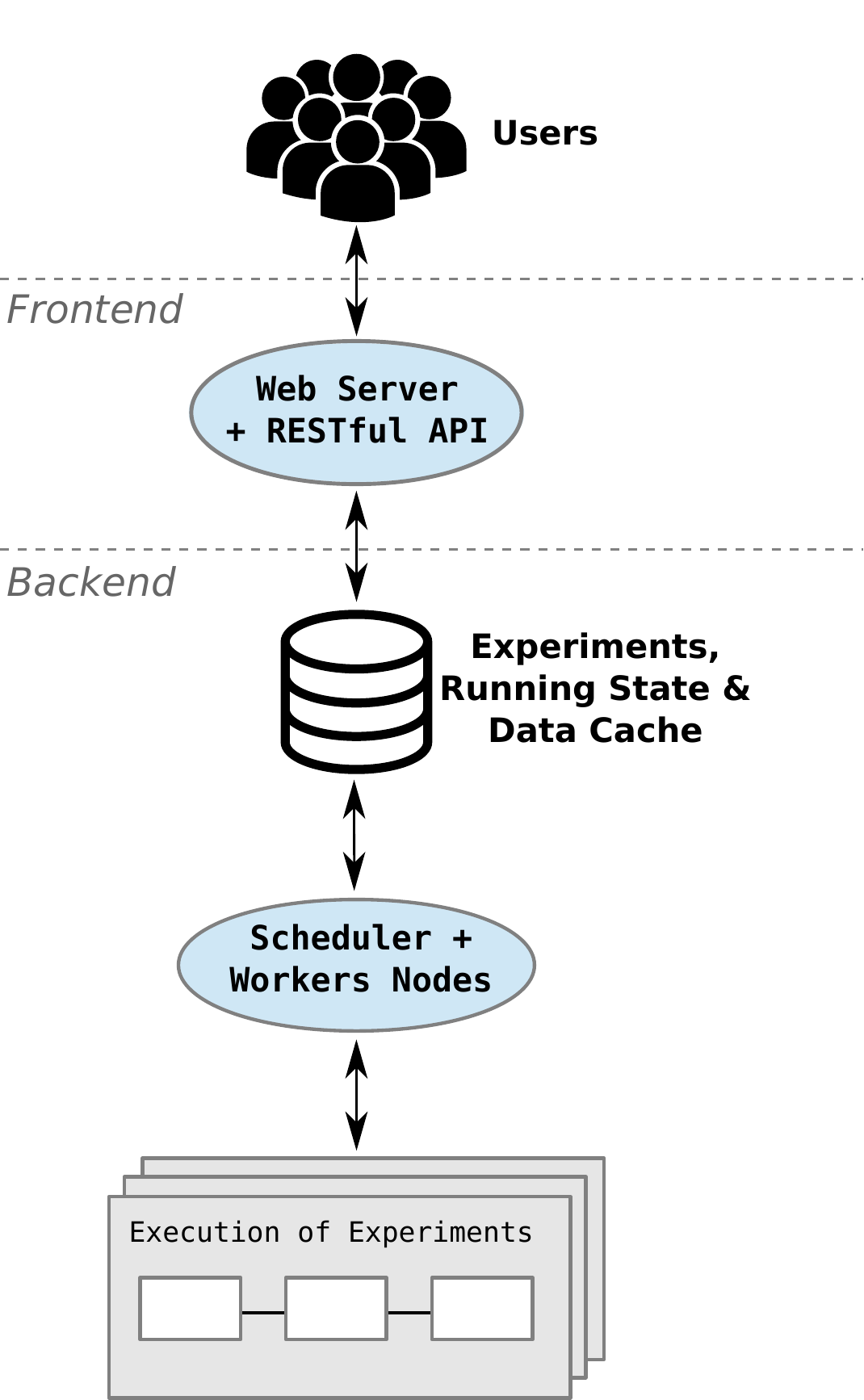}
  \caption{An overview of the BEAT platform applications and their interaction.
  Users use the web front-end to run experiments, search and combine results. A
  back-end handles the execution of experiments on dedicated hardware.}
  \label{fig:beat-high}
\end{figure}

The BEAT Platform is composed of three main applications: the web, the
scheduler and one or more worker nodes. The main function of the web
application is to handle authentication and authorization, while the main
function of the back-end (scheduler and worker nodes) is to handle the
execution of the experiments. Figure \ref{fig:beat-high} shows the high-level
interaction between those three applications.

\begin{sidewaysfigure}
  \begin{center}
    \begin{subfigure}[b]{0.49\textwidth}
      \includegraphics[width=\textwidth]{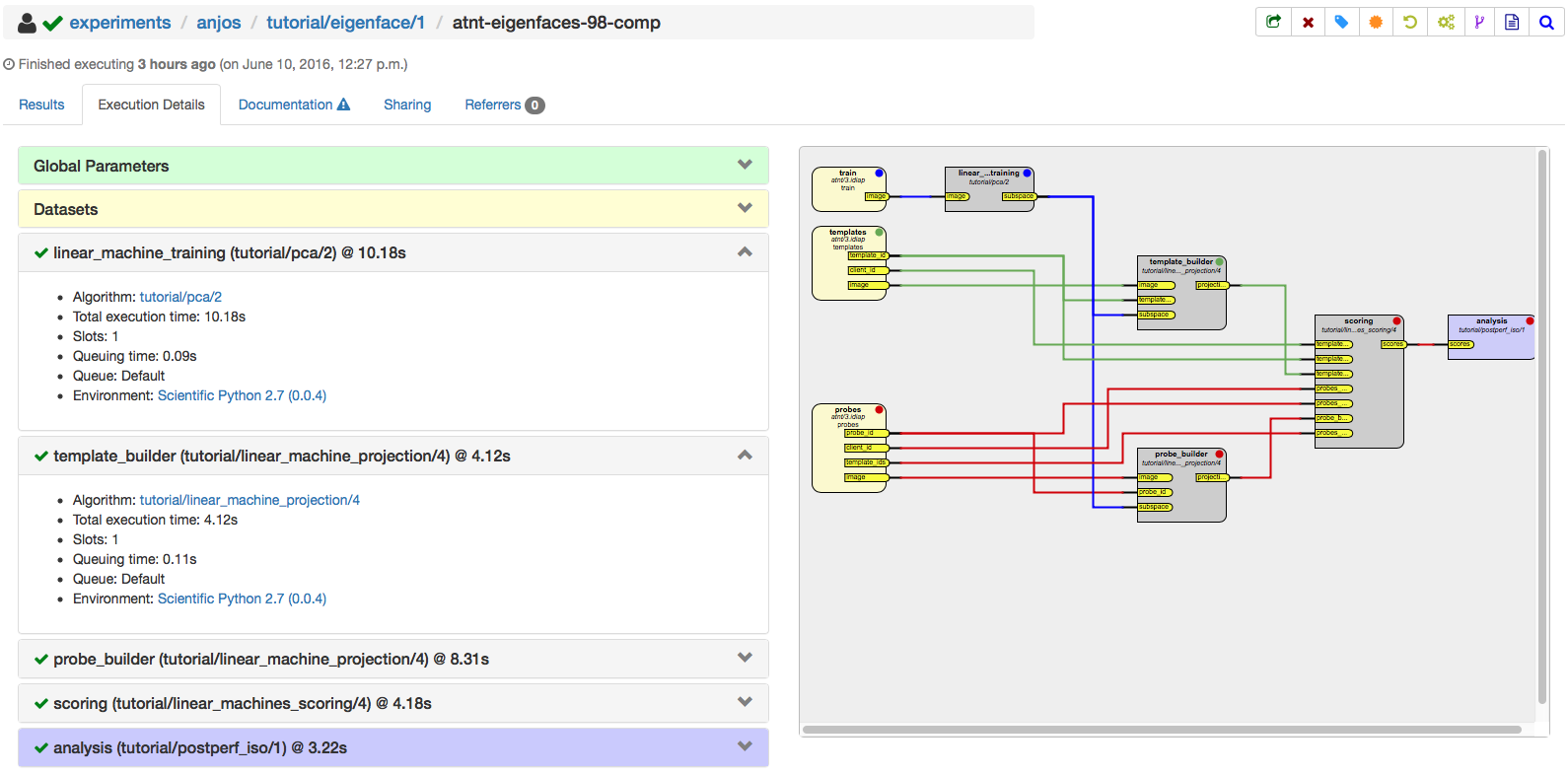}
      \caption{Experiment view}
      \label{fig:experiment}
    \end{subfigure}%
    \begin{subfigure}[b]{0.49\textwidth}
      \includegraphics[width=\textwidth]{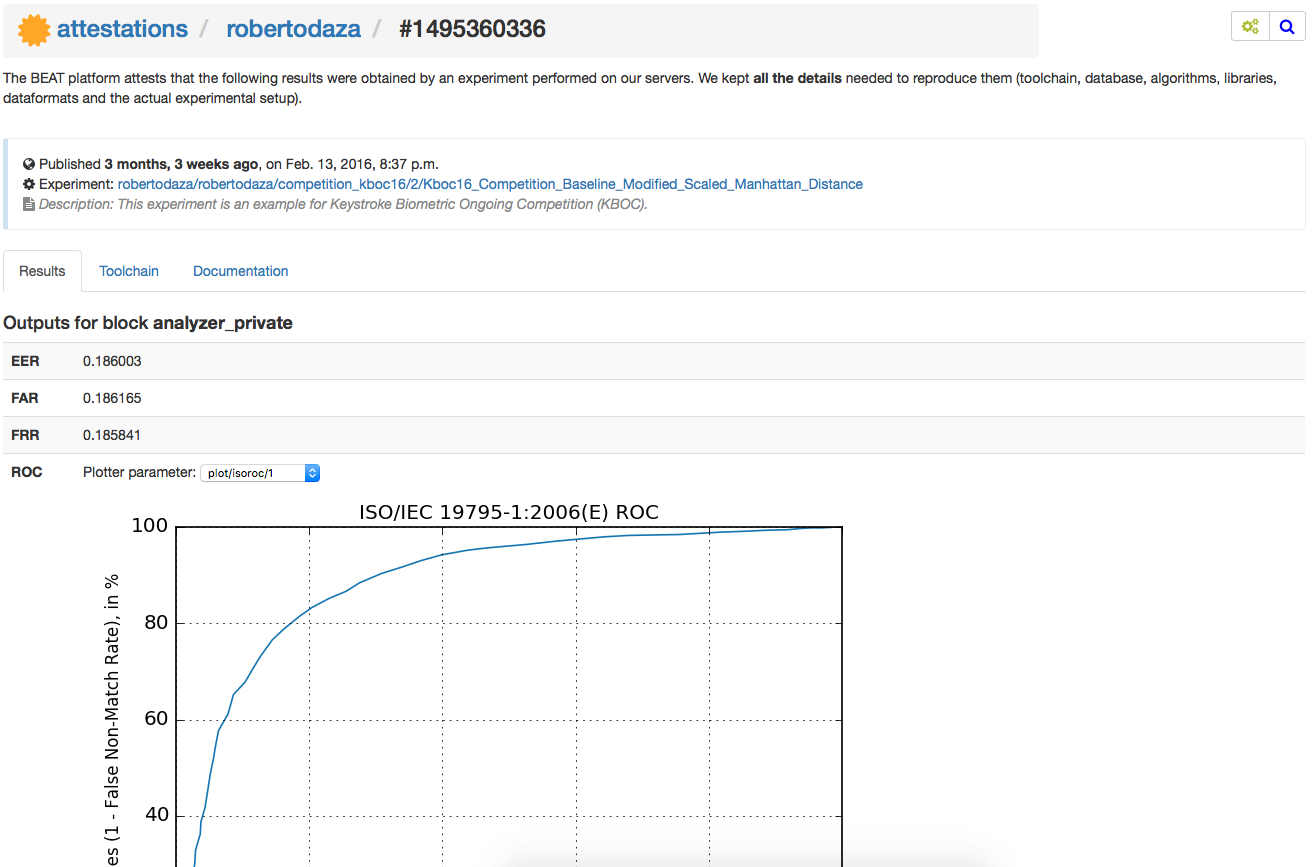}
      \caption{Attestation}
      \label{fig:attestation}
    \end{subfigure}

    \begin{subfigure}[b]{0.49\textwidth}
      \includegraphics[width=\textwidth]{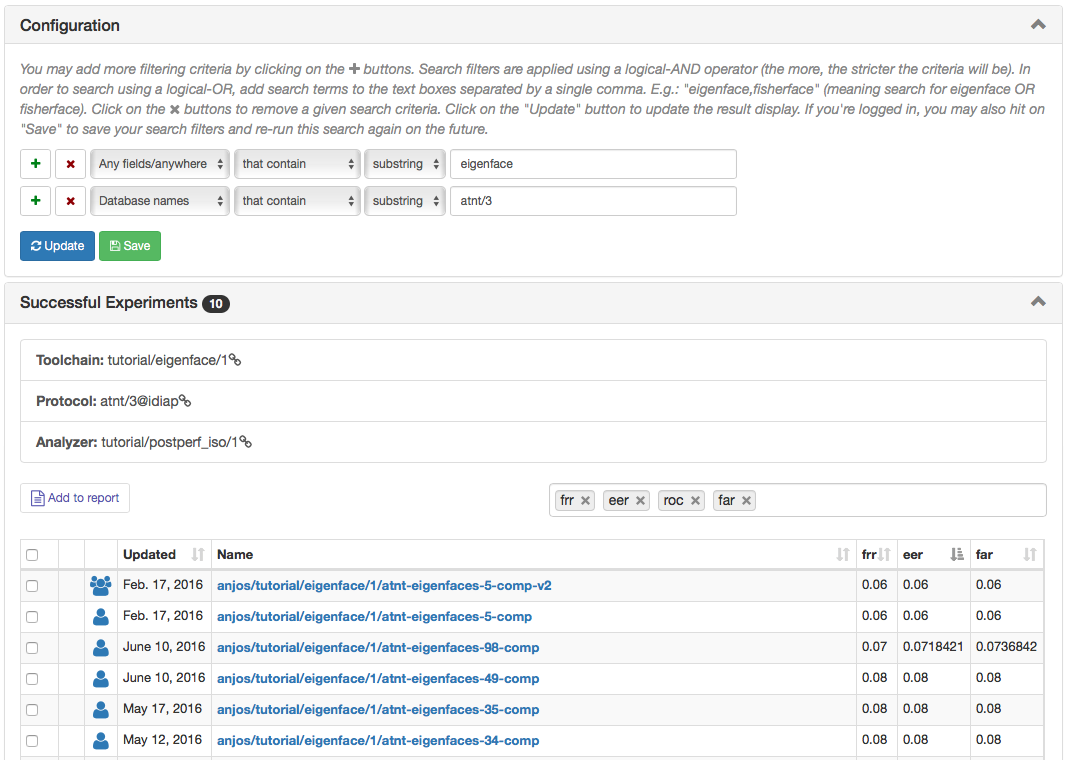}
      \caption{Search view}
      \label{fig:search}
    \end{subfigure}%
    \begin{subfigure}[b]{0.49\textwidth}
      \includegraphics[width=\textwidth]{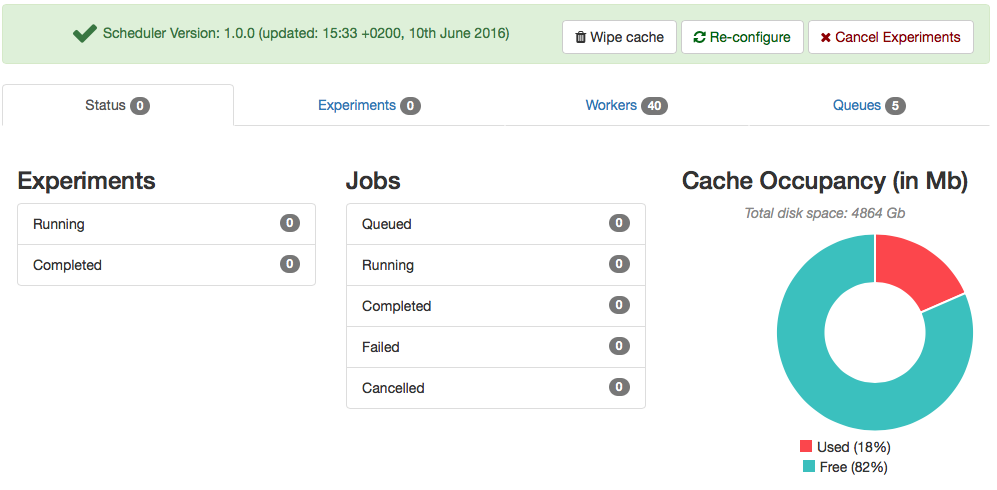}
      \caption{Scheduler control page}
      \label{fig:scheduler}
    \end{subfigure}

  \end{center}

  \caption{BEAT web user interface. (a) Experiment view: after experiment
    execution, the user can check experimental details such as the results, but
    equally computing performance indicators; (b) Attestation: Each experiment
    may be certified by the platform guaranteeing its reproducibility and make
    it read-only; (c) Search and Reports: Experimental results from different
    experiments may be combined into powerful reports that can be exported into
    publications; (d) Scheduler: the backend, composed of workers attached to
    processing queues can be monitored and controlled by platform
    administrators.}
  \label{fig:screenshots}
\end{sidewaysfigure}

The web application is the main entry point of the platform. It consists of two
different major components: a set of \textbf{Web Application Programming
Interfaces (API)} and a \textbf{User Interface}. The BEAT Web API provides a
set of entry points that can be used by external applications that wish to
communicate with the platform, e.g. to get the status of a given experiment.
This Web API is implemented as a RESTful API using the popular Django web
framework~\cite{django}. On top of this API and also written using Django, a
browseable user interface has been built, allowing a user to interact with the
platform via a conventional web browser and to make use of the different
functionalities provided by BEAT, such as implementing algorithms, starting an
experiment or comparing a set of results (see
Figure~\ref{fig:screenshots}). The RESTful API also allows third-party
applications to be developed in order to complement the user experience, for
example, using smart phones or tablets.

The execution of experiments triggered via the web application is performed on
one or more \textbf{workers}, which form the computation back-end. Job
assignment is intermediated by a central \textbf{scheduler} process that
assigns computing jobs to different nodes available respecting user quotas and
hardware requirements. To achieve this flexiblity, the BEAT back-end
closely resembles typical batch-queue submission
systems,\footnote{Batch queuing systems:
\url{https://en.wikipedia.org/wiki/Job_scheduler}} such as the Oracle Grid
Engine\texttrademark or TORQUE and organizes its workers into processing
queues. Each BEAT user can then assign whole experiments to such queues or
individually define which resources each bit of an experiment must use. This
technique allows for experimental toolchains which are composed of an
heterogeneous pool processing environments composed of software libraries and
hardware resources.

\subsection{Object Model}

All interactions between the web and the backend are done using an abstract
object model representing an experiment and associated components, that was
specifically crafted to represent machine learning and pattern recognition
problems. By configuring an experiment, a BEAT platform user puts together a
\textbf{toolchain} (see drawing at
Figure~\ref{fig:experiment}), \textbf{databases} and
\textbf{algorithms} that produce the desired test setup.\footnote{User guide:
\url{https://www.beat-eu.org/platform/static/guide/index.html}}

\begin{figure}[htpb]
  \centering
  \includegraphics[width=\linewidth]{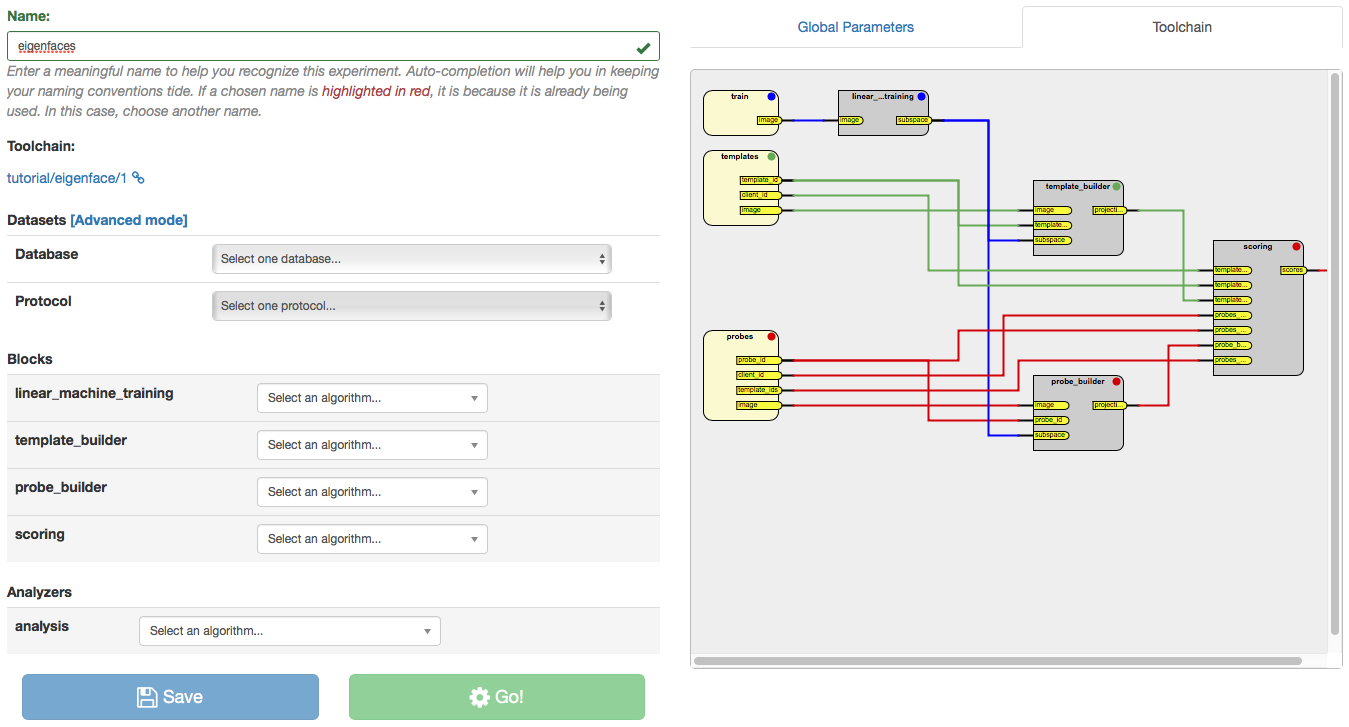}
  \caption{The BEAT platform experiment configurator allows the user to easily
  associate databases, algorithms and analyzers together to create the
  desired setup. As the user chooses components for the toolchain, choices of
  further components are restricted respecting data format compatibility
  between the blocks.}
  \label{fig:configurator}
\end{figure}

A toolchain (or workflow), defines a sequence of interconnected blocks that can
perform a certain task (e.g. face recognition using eigen-faces). Each
connection in a toolchain defines a distinct strongly typed data flow path, and
determines the overall execution order for the blocks. What is not determined
by the toolchain is which algorithms execute in each block or what is the input
database. Once a toolchain is defined, the BEAT platform provides an easy to
use web-based experiment configurator (see Figure~\ref{fig:configurator}) that
allows the user to hand-pick algorithms and databases that fit together
respecting the block configuration, input and output data format exchange
between the block being configured and its surrounding siblings.

\begin{figure}[htpb]
  \centering
  \includegraphics[width=\linewidth]{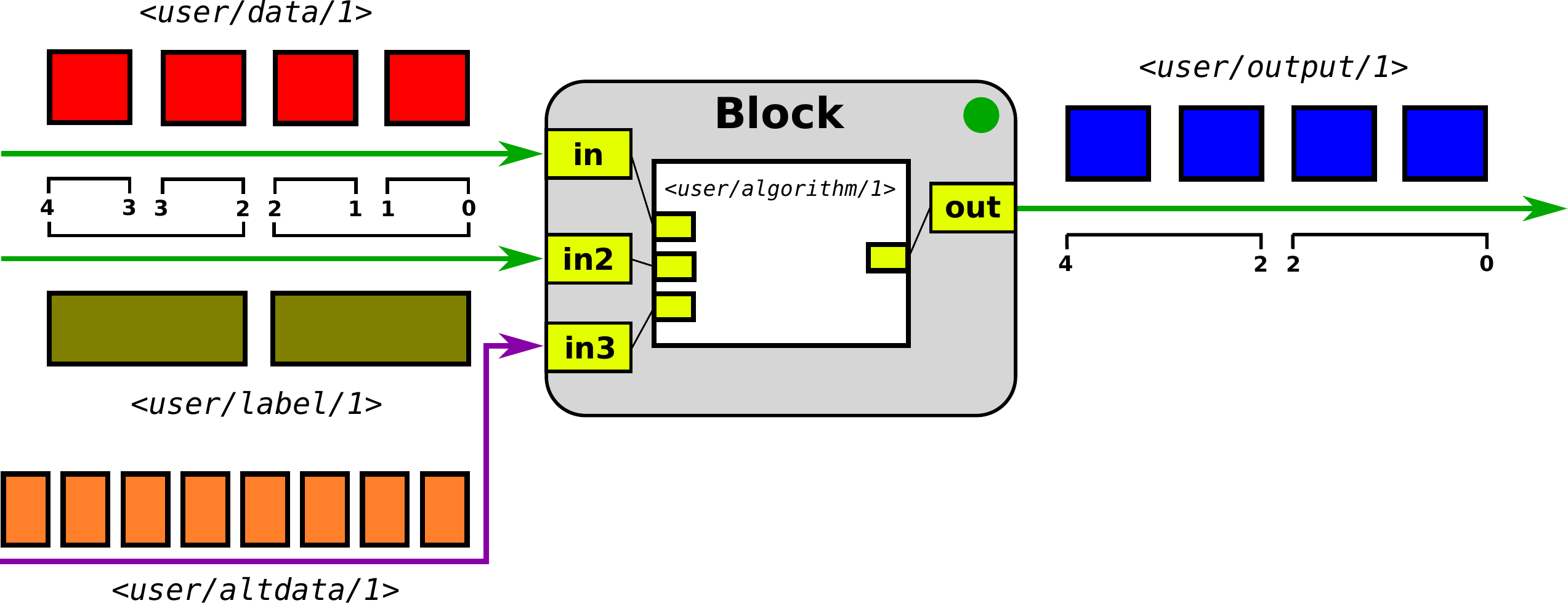}
  \caption{Schematic diagram showing the input of a block, composed of
    different types of objects which are defined by 3 different \textit{data
    formats}, called \texttt{user/data/1}, \texttt{user/label/1} and
    \texttt{user/altadata/1}. Inside the block, an algorithm may be installed
    for as long as it can input those data types. The block fanout data type is
    determined either by the output type of the algorithm which, in turn,
    imposes a restriction on the downstream block algorithms that can be
    installed. Some inputs and the output of a block are automatically
  synchronized together so the BEAT plaform is aware of the relationship
between original data and the outcome of each block in an experiment.}
  \label{fig:block-exchange}
\end{figure}

To implement this feature, the core object model defines \textbf{data
formats}, which are user defined data structures implementing the atomic data
elements that are exchanged in block connections (see more details in
Figure~\ref{fig:block-exchange}). When the user hand-picks a particular
database for the input of an experiment, such a database will yield data
elements of a certain type through its fanouts which limits the choices for the
input blocks containing user code. In the same way, each user algorithm defines
input and output data types, that impose requirements on downstream blocks. The
experiment configurator takes advantage of this feature, coupled with the
algorithms and databases structure (number of fan-ins and fan-outs) to only
provide possible combinations during experiment setup, improving user
experience with the platform.

Each input data block in a toolchain (blocks on the left in
Figure~\ref{fig:configurator}) outputs data through one or more channels that
are synchronized with each other. This technique makes it possible for a
database designer to determine which data points must be served together
respecting data restrictions (e.g. associating labels to samples). Each block
also determines what is the synchronization channel it will work on. Data is
then fed into the algorithm via the block fan-ins in a paced manner, respecting
the original database designer usage protocols. This technique allows the user
code to be programmed loop-free, since iteration is carried out by the platform
itself.

A less obvious yet powerful advantage of this approach is automatic
parallelization. Because the platform controls input data iteration, it is
possible to split processing in N-folds potentially speeding-up data processing
without user intervention.

\subsection{Back-end}

After the experiment is properly configured, it may be stored or scheduled for
execution in the platform processing farm. To address that, a
scheduler application breaks down the execution path and order respecting the
experiment's block boundaries. Each algorithm-equipped block is then executed
in a single isolated process, inputing data from the previous block and
outputing data that is stored into a common, large filesystem based cache. The
BEAT platform cache serves two purposes: (1) it optimizes data processing
allowing scheduled experiments to skip blocks which have already been processed
before, in the context of the same experiments or on identical conditions, and
(2) it isolates inter-block (read process) communication, effectively allowing
for a simpler programming interface while retaining the possibility to execute
each block process in any combination of hardware and software resources
available. In this context, it is possible an experimental setup is effectively
run in a multitude of different environments, e.g., preprocess images using a
python encoded algorithm running on a Linux workstation, but train a deep
convolutional neural network in the next block using a graphical processing
unit (GPU) driven by the Caffe, Torch, or TensorFlow frameworks
\cite{deep-compare:2016}.

To keep resources under control in the back-end processing farm and implement
fair use and sharing, each available compute node (\textbf{worker}) is
allocated to a queue which is subject to usage restrictions. Each queue
determines the maximum memory, number of compute threads and the maximum time
any given user process may consume on any given worker belonging to it. Queues
may be shared among users or groups of users inside the platform giving an
administrative authority the possibility to effectively control resource usage
through the system. When users configure an experiment, they must determine
which processing queues and software environments will be used for each block.
A global default is provided to simplify the configuration process for
the most obvious cases. Any number of user experiments can be scheduled without
limitations. Execution, on the other hand, is subject to resource availability
and priority.

\subsection{Privacy, Certification and Versioning}

The BEAT platform implements a mechanism to control user data confidentiality
and access based on access permissions which are user-configurable. By default,
all user interaction with the platform is kept confidential to users until they
wish to allow access to other parties. The level of control on object access
permissions remotely resembles that of the file systems of UNIX-based operating
systems. Users may allow other users or groups (called \textbf{teams} at the
platform) to view or execute algorithmic code, access experimental results,
toolchains, search queries, data formats, etc. Teams can be created by users
to provide homogeneous access permissions to groups of users to a certain
resource. This feature allows users to share experimental details between
groups of interest (e.g. while building a scientific article), while keeping it
confidential from the general public visiting the platform.

Once experiments were successfuly run, their output (scalars and plots) may be
used in scientific publications. To ensure reproducibility, the BEAT platform
implements a unique attestation mechanism\footnote{US patent filed under the
number US20150970333} that provides an assurance to peer-reviewers and
interested parties all bits making up an experiment (toolchain, algorithms,
data formats and data sets) won't be modified any longer and can be re-used for
verifying results or recombined in new experimental settings. By creating an
attestation, users freeze experiment details altogether (no elements can be
further edited), while allowing anonymous access to most data. Attestation
effectively makes toolchains and data formats used in an experiment public.
Algorithms may be set executable-only which allows peers to re-run the
experiment but no direct access to the source code.

While attested experiments cannot be any longer modified, it is still possible
for users to create new versions of individual components by copying
existing material available. New versions are tracked by the platform, which
records code re-use and can therefore potentially be used to trace original
authorship.

\paragraph{Databases:} One important aspect on data privacy may concern raw
input data, that is finally fed by the BEAT platform backend into experiments.
Databases in sensitive areas such as those related to biometrics, forensics or
biomedical applications, may require special end-user license agreements that
require data in its original or processed forms is not exported from the peer
institute premises.

To comply with such requirements, the BEAT platform is designed taking
privacy as a very strong and important operational constraint. Only platform
administrators are allowed to physically copy raw data sets into the
platform file systems. Once such data sets are in place, administrators create
special algorithms that explain how to read data from disk and feed it into
user toolchains while respecting usage protocols defined by the original data
controller. Such special algorithms are called \textbf{views} and behave in a
similar way to user algorithms. Users are allowed to plug in one or more
database views into experiments effectively defining the input blocks of the
experiment. Because intermediate data output by processing blocks cannot be
exported, data privacy is guaranteed.

\section{Applications}
\label{sec:use}

In its current state, the BEAT platform can honour a large number of use cases
in pattern recognition and machine learning applications through the use of its
built-in experiment running facilities, certification mechanisms and storable
searches terms. In this section, we investigate a few of these use-cases and
how BEAT provides an answer to hosting challenges (competitions) or can be made
useful in industry-academia relationships.

\subsection{Use-case 1: Cooperative academic work}

Development of new techniques and ideas in academic work very often is the
result of team work. Two colleagues studying a particular subject may be on the
same academic premises or apart, in completely different institutions. Ideas
are exchanged via e-mails and virtual meetings, a common software framework is
selected as a method for sharing implementations and analysis is carried out by
the exchange of simple scalars, tables and, frequently, graphics comparing
results of experimental setups. Go/no-go decisions are taken in group as part
of the analysis process, which typically involves an even greater number of
persons (e.g. research supervisors). The BEAT platform can be used in this
context to ensure that the same software environment and analysis is executed
for all experiments leading to a scientific report, guaranteeing homogenity and
reproducibility at all times.

After databases are installed into the platform, one of the parties in the
academic team creates (or copies) algorithms, toolchains and experiments that
represent the state-of-the-art baseline systems one wishes to improve upon. At
this point a clear set of metrics is defined leading to scalars and figures
which should be generated on a per-experiment basis. These elements are then
shared via the BEAT team feature so that only people in the academic pool of
interest have access to core elements of the study and end results.

\begin{figure}[htpb]
  \centering
  \includegraphics[width=\linewidth]{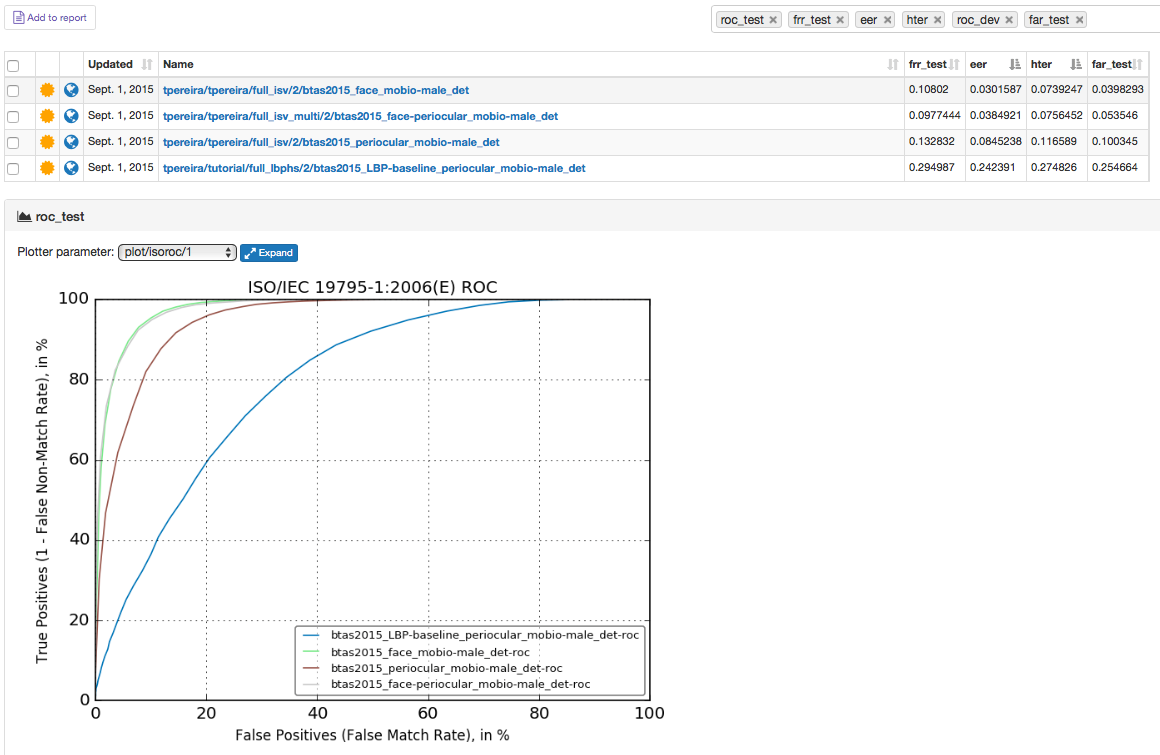}
  \caption{Example of tables automatically generated by the BEAT platform via
    its \textit{search} feature. It allows you to store specific search terms
    that result in scalars and figures from the selected experiments to be
    compared. The user can select how to display the figures associated with
    their analysis metrics, how to sort and which columns to show on the
  comparison table. Available online via
\url{https://www.beat-eu.org/platform/search/tpereira/btas2015_mobio_male/}.
The resulting report can be found at
\url{https://www.beat-eu.org/platform/reports/751803513/} and provides
assurance experiments in the paper \cite{btas:2015} are reproducible.}
  \label{fig:use-case-1}
\end{figure}

A search term is then created by a team member (and shared) such that it is
possible to keep track of advances when comparing various experiments together.
Figure~\ref{fig:use-case-1} shows an example search filter setup on the BEAT
platorm, examplifying a possible setup. The saved search query can be completed
with a rich text description. Users have the ability to control which analysis
figures to display in the aggregation table and plots. An unlimited number
search terms may be stored to express different analysis points of
views. For example, it would be possible to setup a search query that would
compare the user current work against other algorithmic approaches or, in
another instance, how the new user setup improves across databases or different
parameter sets.

Once a conclusion has been reached and a scientific report needs to be
generated, it is possible to attest all experiments together through BEAT
platform \textit{reports}. A report can be thought as a
\textit{macro-attestation}, certifying multiple experiments together in a
single place. Reports can also be decorated with documentation, can contain
tables and figures and be locked for external peer-reviewing, giving an
assurance of reproducibility to external reviewers and the research team.

A few articles have already been published using the platform \cite{btas:2015,
korshunov:2016, kboc:2016}, demonstrating its capabilities to serve as a
reproducible research tool and a portal for extensible and re-usable
experimental code and documentation.

\subsection{Use-case 2: Challenges and education}

It is a common practice in statistical pattern recognition research domains to
setup challenges in order advancing research for specific goals. For example,
in the area of Biometrics, most large conferences contain specific sessions for
competitions organized to, for instance, improve the state-of-the-art in
biometric authentication, identification, \textit{spoofing} detection or
pre-processing techniques. An academic or industrial partner sets up the
competition defining one (or more) databases that will be used for the
challenge, together with usage rules that must be respected by all
participants. In common practice, final evaluation samples are normally not
provided from the begin of the competition, to avoid bias in the final results.

The BEAT platform is a challenge-ready system requiring minimal management
intervention during the course of the competition, from the part of the
organizer. After databases and usage protocols are properly installed, the
competition organizer may define baseline toolchains, algorithms and an
analysis metric that will be provided to participants. A search query may then
be stored which may define the criteria that must be respected in the
competition. For instance, participants may be required to use a particular
toolchain, algorithm, database version and/or protocol in their submissions.
As the competition advances, the competition organizer may resubmit the store
search query that will display an updated version of the leaderboard. The last
step can be automated by the platform (via a specific \textit{leaderboard}
checkbox) and, in this case, competition organizers will receive an e-mail each
time the search results for a stored search query changes. Competitors, when
ready, share back their algorithm implementations with the organizer, that then
can visually compare results through their stored search query. By design,
entries originating from violations of the established competition rules simply
do not appear in the defined leaderboard. The recently organized Keystroke
Biometrics Ongoing Competition (KBOC) is an example of challenge using the
platform.\footnote{\url{https://www.beat-eu.org/platform/search/aythamimm/KBOC16_COMPETITION_SEARCH/}}
Final competition results were published in \cite{kboc:2016}.

The BEAT platform may be equally used as an academic resource in a
teacher-student relationship, in ways very similar to the organization of a
competition. In this use-case, teachers define a task using the BEAT platform
and may ask students to fill-up algorithmic details for exercising or for
grading. As students finish their assigment, they can share back their work
with the teacher using the specific sharing features of the platform.

\subsection{Use-case 3: Industry-Academia interchange}

Industry and academia often work on similar research topics. More specifically,
an industrial partner may be interested in evaluating a technique developed by
researchers in academia, before considering its potential exploitation.

Unfortunately, there may be several practical issues preventing a fruitful
collaboration. First, the industrial world is always eager for incremental
deliverables, before making the decision of pursuing or stopping a
collaboration. This is not always easy when the company and its academic
partner have completely different engineering processes. Second, a company may
want to know how much better a system is before negotiating a license for its
commercial exploitation. However, the company may be unable to share a dataset
because of legal and/or privacy concerns. Third, technology transfer is often a
difficult process where discrepancies may emerge due to misunderstandings on
how to evaluate a technology and on which metrics and evaluation protocol to
use on a given dataset. Finally, it may be difficult for an industrial partner
to modify and re-evaluate a delivered prototype on another dataset, since an
employee has to figure out how the prototype can be used and how to evaluate it
on a given dataset. At the end, all those difficulties may negatively affect
the decision of a company to engage with an academic partner.

The BEAT platform provides several features that address many of the above
mentioned issues. Typically, the industrial partner would first set up a BEAT
platform instance installed within its premises. Then, the database for the
project is setup, and well defined evaluation protocols are implemented after
a joint agreement between both parties. All of this successfully addresses the
issue of giving an access to evaluate a prototype on this database to external
partners, while keeping the data sequestered to respect legal and privacy
concerns. The next step for the industrial partner is to implement a baseline
system on the platform, allowing the academic partner to better understand
what kind of prototype is expecting from him, as well as one (or several)
analyzers that implement metrics and the visualization tools required to
evaluate a system. Besides, the industrial partner may clearly indicate to the
other party which components they are allowed to change on the system. It can
be an algorithm on a given box or the whole toolchain and associated algorithms
depending on the project requirements.  Analysis should be preserved as that
corresponds to the metric of interest to the industrial partner. The academic
partner then starts to upload prototypes until the project ends or the goal is
satisfied. Benchmarks tell the academic partner how much close to the goal they
are. Industrial partner gets a working, reproducible prototype at all times.

\section{Conclusions}
\label{sec:conclusions}

The BEAT platform is an open-source software platform for data researchers and
data owners, which allows executing and evaluating algorithms on image, audio,
video, or multi-dimensional data sets. It can host data that cannot be
distributed by conventional means, either because of their large size or
because of confidentiality constraints (or both). It offers data-owners the
possibility (in agreement with the researchers) to select the processing
pipelines most appropriate for their needs, while offering researchers access
to big data while minimizing the legal hassles, risks, and cost that accompany
conventional data sharing and that currently hamper the research community to
fully contribute to solving challenges associated to big data. The software
stack of the platform is composed of nearly ~80'000 lines of code (Python:
58\%, Javascript: 24\%, HTML: 15\%, CSS: 3\%).

Despite the large number of features, the BEAT platform also presents
limitations and therefore a lot of potential for improvements. Work leading to
a more useful platform still needs to address the following issues:

\begin{itemize}
  \item Multi-site scalability: Platforms running on different sites cannot
    currently communicate between each other. Effectively, it is both
    impractical and expensive to keep copies of large datasets in every
    institution where research is being conducted. It implies redundant
    expenses and may lead to reproducibility issues due to incomplete datasets
    existing in various places as a result of lack of resources or human
    errors;

  \item More backends: As of today, only a Python processing backend is
    implemented and available on the platform, allowing user algorithms to be
    programmed on. This may be considered an important limitation for different
    research labs in which other programming backends are used (e.g. Matlab, R,
    Julia, privately compiled binaries, etc.). A possible model to be adopted
    in this case is that of running user code in containers or virtual
    machines, which are isolated from the BEAT infrastructure. With such an
    approach, one may introduce flexibility while also enforcing security
    constraints;

  \item Development and debugging of experiments: Finally, current provisions
    for developing and debugging code for the BEAT platform could be improved.
    The migration from a desktop typical research environment into the
    web-based system may prove difficult for inexperienced users. A system to
    export and run BEAT experiments on someone's computer is available.
    Unfortunately, it requires users install raw data and a compatible software
    stack to make it functional, which can be considered a complex task. Work
    in this context should try to reduce or completely remove this entrance
    barrier by providing a development environment where experiments can be
    executed w/o requiring the user to manage software stacks while still
    allowing full control of the code flow in order to craft new components or
    debug existing ones.

\end{itemize}

We continue to develop the platform towards these goals and welcome new groups
and partnerships to enlarge the domains of application of our currently
operational prototype.

\section{Acknowledgements}

The research and development required by the platform and leading to this
article has received funding from the European Community's FP7 under the grant
agreement 284989 (BEAT) and from the Swiss Center for Biometrics Research and
Testing (\url{www.biometrics-center.ch}).

The authors would like to thank also all the talentous engineers from the Idiap
research institute: Philip Abbet, Samuel Gaist, Flavio Tarsetti, Frank Formaz
and Cédric Dufour. Without their help, an operational platform would not have
been possible.

\bibliographystyle{ieeetr}
\bibliography{references}

\begin{thebibliography}{10}

\bibitem{price:1986}
K.~Price, ``Anything you can do, {I} can do better ({N}o you can't)...,'' {\em
  Computer Vision, Graphics, and Image Processing}, vol.~36, pp.~387--391,
  1986.

\bibitem{nature:2016}
M.~Baker, ``1,500 scientists lift the lid on reproducibility,'' {\em Nature:
  International Weekly Journal of Science}, vol.~533, May 2016.

\bibitem{jasny:2011}
B.~R. Jasny, G.~Chin, L.~Chong, and S.~Vignieri, ``Again, and again, and
  again...,'' {\em Science}, vol.~334, no.~6060, p.~1225, 2011.

\bibitem{vandewalle:2012}
P.~Vandewalle, ``Code sharing is associated with research impact in image
  processing,'' {\em IEEE Computing in Science and Engineering}, vol.~14,
  pp.~42--47, July 2012.

\bibitem{executable-paper}
``Executable paper grand challenge.'' \url{http://www.executablepapers.com/}.
\newblock Accessed: 2016-06-10.

\bibitem{mit-generator}
``Scigen.'' \url{https://pdos.csail.mit.edu/archive/scigen/}.
\newblock Accessed: 2016-02-11.

\bibitem{codalab}
``Codalab: Accelerating reproducible computational science.''
  \url{http://codalab.org/}.
\newblock Accessed: 2015-12-14.

\bibitem{kaggle}
``Kaggle: The home of data science.'' \url{https://www.kaggle.com/}.
\newblock Accessed: 2015-12-14.

\bibitem{eu-law-reform}
``Reform of eu data protection rules.''
  \url{http://ec.europa.eu/justice/data-protection/reform/index_en.htm}.
\newblock Accessed: 2016-06-10.

\bibitem{django}
``Django: The web framework for perfectionists with deadlines.''
  \url{https://www.djangoproject.com}.
\newblock Accessed: 2016-06-10.

\bibitem{deep-compare:2016}
S.~Bahrampour, N.~Ramakrishnan, L.~Schott, and M.~Shah, ``Comparative study of
  caffe, neon, theano, and torch for deep learning,'' {\em CoRR},
  vol.~abs/1511.06435, 2016.

\bibitem{btas:2015}
T.~de~Freitas~Pereira and S.~Marcel, ``Periocular biometrics in mobile
  environment,'' in {\em IEEE Seventh International Conference on Biometrics:
  Theory, Applications and Systems}, pp.~1--7, IEEE, Sept. 2015.

\bibitem{korshunov:2016}
P.~Korshunov and S.~Marcel, ``Joint operation of voice biometrics and
  presentation attack detection,'' in {\em IEEE International Conference on
  Biometrics: Theory, Applications and Systems (BTAS)}, (Niagara Falls, NY,
  USA), pp.~1--6, Sept. 2016.

\bibitem{kboc:2016}
A.~Morales, J.~Fierrez, M.~Gomez-Barrero, J.~Ortega-Garcia, R.~Daza, J.~V.
  Monaco, J.~M. {a}o, J.~Canuto, and A.~George, ``Kboc: Keystroke biometrics
  ongoing competition,'' in {\em 8th IEEE International Conference on
  Biometrics: Theory, Applications, and Systems}, 2016.

\end{thebibliography}
\end{document}